\documentclass[aps,prd,preprint,superscriptaddress,nofootinbib,11pt]{revtex4-1}
\usepackage{graphicx,subfigure}
\usepackage{amsmath,amssymb,amsthm,mathtools}
\usepackage{geometry}
\usepackage{appendix}
\usepackage[colorlinks=true, allcolors=blue]{hyperref}
\usepackage{bm}

\graphicspath{{figures/}}
\begin{document}

\title{Post-Newtonian Binary Dynamics in Effective Field Theory of Horndeski Gravity}
\author{Wen-Hao Wu}
\email{wuwenhao21@mails.ucas.ac.cn}
\affiliation{School of Astronomy and Space Science, University of Chinese Academy of Sciences (UCAS), Beijing 100049, China}
\author{Yong Tang}
\email{tangy@ucas.ac.cn}
\affiliation{School of Astronomy and Space Science, University of Chinese Academy of Sciences (UCAS), Beijing 100049, China}
\affiliation{School of Fundamental Physics and Mathematical Sciences, Hangzhou Institute for Advanced Study, UCAS, Hangzhou 310024, China}
\affiliation{International Center for Theoretical Physics Asia-Pacific, Beijing 100049, China}

\begin{abstract}
General relativity has been very successful since its proposal more than a century ago. However, various cosmological observations and theoretical consistency still motivate us to explore extended gravity theories. Horndeski gravity stands out as one attractive theory by introducing only one scalar field. Here we formulate the post-Newtonian effective field theory of Horndeski gravity and investigate the conservative dynamics of the inspiral compact binary systems. We calculate the leading effective Lagrangian for a compact binary and obtain the periastron advance per period. In particular, we apply our analytical calculation to two binary systems, PSR B 1534+12 and PSR J0737-3039, and constrain the relevant model parameters. The theoretical framework can also be extended to higher order systematically. 
\end{abstract}

\maketitle

\section{Introduction}
General relativity (GR) has been the most successful theory of gravitation since its proposal by Einstein. It has been tested by various precision experiments and provides an essential theoretical framework for cosmology. However, there are still several cosmological puzzles that GR cannot provide satisfactory explanations for, including the observed tiny cosmological constant or the dark energy, the nature of dark matter, the beginning of our universe, etc. Partly motivated by these observational puzzles and theoretical issues, various modified gravity theories have been proposed and offered alternative frameworks for the currently well-accepted $\Lambda$CDM paradigm.

%  Model: Horndeski Gravity

One of the simple ways to modify GR is to add new degree of freedom~\cite{Odintsov:2023weg,Clifton:2011jh, Sotiriou:2008rp, DeFelice:2010aj, Nojiri:2017ncd, Cai:2015emx, deRham:2014zqa}, for instance, the Jordan-Brans-Dicke scalar-tensor theory~\cite{Brans:1961sx}, k-essence theory~\cite{Armendariz-Picon:2000ulo,Armendariz-Picon:1999hyi,Armendariz-Picon:2000nqq}, Galileon theory~\cite{Nicolis:2008in, Deffayet:2009mn, Deffayet:2011gz}, gauge theories of gravity~\cite{Wu:2015wwa,Wu:2017urh,Wu:2022mzr,Wu:2022aet}, Starobinsky model~\cite{Starobinsky:1980te} and its extension with Weyl symmetry~\cite{Ferreira:2019zzx, Ghilencea:2019rqj, Tang:2020ovf}, etc. To be theoretically consistent, such models should satisfy well-definedness conditions. For example, the equation of motion would be of second order in derivatives to avoid instabilities or so-called ``ghosts"~\cite{Ostrogradsky:1850fid}. With these considerations, Horndeski gravity stands out as a well-formulated and highly general one~\cite{Horndeski:1974wa, Charmousis:2011bf, Kobayashi:2019hrl}. It is constructed by simply adding one scalar degree of freedom to GR. Therefore, it serves as a good starting point for exploring the parameter space of modified gravity theories~\cite{Gao:2014soa, Koyama:2015vza, Kase:2018aps, Kobayashi:2019hrl, Quartin:2023tpl}.  

% binary pulsar system

The observation of gravitational waves (GW) from compact binaries since 2015 has provided a new platform for the test of gravitation theories~\cite{LIGOScientific:2017vwq}. For example, the merger of neutron stars GW170817~\cite{GW170817} has shown that the speed of GW is very close to the speed of light~\cite{LIGOScientific:2017zic}. In general, compact binaries are canonical objects for gravitating systems. Besides the GW signals, the emission of electromagnetic pulse signals has provided useful information about their orbital dynamics as well, and such binary pulsars can be used to test gravity theories~\cite{Hu:2023vsq, Avdeev:2018ihq}. For our purpose, we take the precession of the orbital motion to test Horndeski gravity due to the relatively high precision. We compare the theoretical predictions with the observations for two binary pulsar systems, PSR B 1534+12~\cite{Fonseca:2014qla} and PSR J0737-3039~\cite{Lyne:2004cj}. Formulating the post-Newtonian (PN) dynamics of the binary system in Horndeski gravity, we provide a self-contained approach to constrain the parameter space of such a theory. 

% The PN-EFT Approach

Concretely, we investigate the dynamics of inspiral compact binary systems in Horndeski gravity. We extend a PN effective field theoretical (EFT) approach in GR~\cite{Goldberger:2004jt, Goldberger:2006bd, Goldberger:2007hy,Bhattacharyya_2023} in the perturbative regime. By taking the orbital velocity as an expansion parameter, we calculate the leading contributions and integrate out the potential mode to get an effective action for the binary system. This formulation gives us a systematic way to evaluate the effective dynamics of any order of orbital velocity analytically. In particular, we are able to include the contribution from the self-interaction of the scalar field and give a bound for the coupling constant from the observation of binary pulsar. This approach can be extended to include higher-order dynamics systematically.

% The Division of the Article

This paper is organized as follows. In section~\ref{sec:model}, we introduce the formalism of Horndeski gravity for a gravitating compact binary and build up the full Lagrangian. Then in section~\ref{sec:eft}, we introduce the post-Newtonian EFT approach for calculating the binary dynamics and give the detailed Feynman rules from the expansion. In section~\ref{sec:PN}, we obtain the orbital precession in conservative dynamics of a binary at first PN (1PN) order by calculating a series of the Feynman diagrams needed for the 2-body dynamics in Horndeski gravity. In section~\ref{sec:numeric}, we compare our results with the observations of binary pulsars and constrain the model parameters. Finally, we summarize the results and give our conclusion.

Throughtout our discussion, we use mostly-minus signature $(+,-,-,-)$ for the metric $\eta_{\mu\nu}$.
    
\section{Horndeski Gravity}\label{sec:model}

Horndeski theory is the class of modified theories of gravity that introduce an extra scalar degree of freedom $\phi$ while having a second-order equation of motion for the fields. Thus, it avoids the Ostrogradski instability caused by higher derivatives. Its action has the general form,
\begin{align}
 S & \supset S_{EH} +  \int _x  \Bigl\{  X + G_2(\phi,X)+ G_3(\phi,X)\square \phi + G_4(\phi,X) R + G_{4,X} \left[ (\Box \phi)^2 - \phi_{\mu\nu} \phi^{\mu\nu} \right]   \nonumber \\
 &  + G_5(\phi,X) G^{\mu\nu} \phi_{\mu\nu} - \frac{G_{5,X}}{6} X \left[ (\Box \phi)^3 - 3 \Box \phi \phi_{\mu\nu} \phi^{\mu\nu} + 2 \phi_{\mu\nu} \phi^{\nu\lambda} \phi^{\mu}_{\lambda} \right]\Bigr\},
\end{align}
 where the spacetime integration is defined as $\int _x\equiv \int \mathrm{d}^4 x \sqrt{-g}$, $g$ is the determinant of metric field $g_{\mu\nu}$, $S_{EH}=-2 m_p^2 \int_x R$ is the Einstein-Hilbert action of Ricci scalar $R$ determined by $g_{\mu\nu}$, $m^{-2}_{p} = 32\pi G$, $G$ is related to Newton's constant, $G^{\mu\nu}$ is the Einstein tensor and $G_n(\phi,X)$ are functions of $\phi$ and $X$, with notations 
 \[  
 X=\frac12g^{\mu\nu}\nabla \phi_{\mu} \nabla_{\nu} \phi, \quad \phi_{\mu_1,\ldots, \mu_n} = \nabla_{\mu_1} \ldots \nabla_{\mu_n}\phi, \quad f_{,A} =\partial_{A}f \; \text{\,for any function $f$ of $A$}.
 \] 

We impose that the speed of GW propagation is the same as light, then only $G_2(\phi,X)$, $G_3(\phi,X)$ and $G_4(\phi)$ remain~\cite{Kobayashi:2019hrl}. To calculate the leading PN effective Lagrangian, we only need to expand the full Lagrangian for 3-point vertices. In addition, linear $\phi$ or $X$ terms in $G_2$ can be shifted away by redefinition of $\phi$. Furthermore, we consider massless scalar so that a $\phi^2$ term is not included here. And a 3-point vertex $g_2 \phi X$ term can be combined with the kinematic term of $X$ to redefine $\sqrt{1+g_2 \phi}\,d\phi \to d\Phi$ to normalize kinetic term canonically. Therefore, the leading term in $G_2(\phi,X)$ for our interest is $\phi^3$. For a detailed discussion about the removal of $g_2 \phi X$ term, see Appendix \ref{sec:app1}.

Let us consider the following action,
 \begin{equation}
 S \supset S_{EH} + \int_x \Bigl\{ X + G_2(\phi,X)+ G_3(\phi,X)\square \phi + G_4(\phi) R \Bigr\}.
 \end{equation}
In particular, we consider the following case for illustration of the dynamics,
\begin{equation}\label{eq:parameters}
    G_2 = g_2 m_p\phi^3/6,\quad G_3 = g_3m_p^{-3} X, \quad G_4 = 2 g_4 m_p \phi, %g_3 m_p^{-3} \,X ,
\end{equation}
where $g_2, g_3$ and $g_4$ are normalized dimensionless constants that are to be determined or constrained by observations. Note that $g_3$ is parameterized in such a way so that it is dimensionless, by introducing the inverse of Planck mass which is large comparing to energy scales of the typical process in a binary pulsar system. This suggests that the viable $g_3$ can be a big number, still allowed by the bound of from binary pulsar, as we shall see in later discussions. 

In addition, we should add the gauge-fixing (GF) term associated with the gauge invariant action $S_{EH}$, $S_{GF}$, which in harmonic gauge is given by
\begin{align}
   S_{GF} &=  m_p^2 \int _x \,g_{\mu\nu} \Gamma^{\mu} \Gamma^{\nu}, \quad \Gamma^{\mu} = g^{\rho\sigma} \Gamma^{\mu}_{\enskip\rho\sigma},
\end{align}
where $\Gamma^{\mu}_{\enskip\rho\sigma}$ is the connection determined by the metric $g_{\mu\nu}$.

As we only consider the inspiral phase of the compact binary in which the stars are moving much slower than light, we can treat the binary system classically. Each star of the binary system is regarded as a point particle with mass $m_a$,  locating at $x^{\mu}_a(t)$ and interacting with the gravitational field, which is described by a worldline Lagrangian $S_{pp}$,
\begin{equation}
    S_{pp} = -\sum_{a=1,2} m_a \int \mathrm{d}\tau_a 
    = -\sum_{a=1,2} m_a \int \mathrm{d}t \sqrt{g_{\mu\nu}(x_a(t)) \frac{\mathrm{d}x_a^{\mu}}{\mathrm{d}t}\frac{\mathrm{d}x_a^{\nu}}{\mathrm{\mathrm{d}}t}}.
\end{equation}
To summarize, the full action for our later discussion is
\begin{align}\label{eq:lag}
S = \int_x \left[ m_p^2 \left( g_{\mu\nu} \Gamma^{\mu} \Gamma^{\nu} - 2R\right)
        +  X  + \frac{g_2}{6} m_p\phi^3 +  \frac{g_3}{m_p^{3}} X \square \phi - \, 2 g_4 m_p \, \phi R \right] - \sum_{a=1,2} \int m_a \mathrm{d} \tau_a .
\end{align}
We can observe that the scalar field does not couple to worldline Lagrangian (matter) directly, but only indirectly through gravity, which shall simplify the calculations, unlike previous investigations on scalar-tensor theories in which the scalar field couples to matter directly~\cite{Huang:2018pbu, Kuntz:2019zef, Brax:2021qqo, Diedrichs:2023foj}.

\section{Post-Newtonian Expansion and Effective Field Theory}\label{sec:eft}

To work out Horndeski theory for a compact binary system in the inspiral phase, we use an effective field theoretical approach based on PN expansion~\cite{Goldberger:2004jt}, which is often also called the non-relativistic GR (NRGR) method. In the inspiral phase of a compact binary system, the orbital velocity $v$ is in non-relativistic regime and much less than the speed of light. Hence, it is meaningful to separate the scales of energy with respect to the post-Newtonian expansion parameters $v$. In particular, we have 3 relevant physical scales in the system: the Schwarzschild radius of of system, $r_s = 2 G m$, the radius of the orbit, $r$, and the wavelength of gravitational wave, $\lambda$. 

In a non-relativistic regime, the system satisfies the virial theorem $ v^2 \sim Gm/r$, thus we have $r_s/r \sim v^2 \ll 1$. While from $\lambda \sim  r/v$, we have $r/\lambda \sim v \ll 1$. So these scales are separated in the non-relativistic limit $v \ll 1$. The physics at the scale $r_s$ is only relevant in strong field regime and contained in the effective mass of point particles. The main physics in the system is carried by gravitons with 4 wave-vector of order
$(\omega, |\mathbf{k}|) \sim (v/r,1/r)$ in potential region and of order $(\omega, |\mathbf{k}|) \sim (v/r, v/r)$ in radiation region. Furthermore, by integrating the potential gravitons, we will get the effective potential and the effective coupling of radiation to the systems.

To perform the calculations in the effective field theory, we have to expand a field in different modes of different energy scales. For Horndeski theory, apart from the metric field whose expansion yields two tensor modes of gravitons, we will have an extra scalar field. Both the scalar and tensor degrees of freedom have their potential and radiation modes. Expanding the metric tensor and scalar fields in different regions, we have
\begin{equation}
 g_{\mu\nu} - \eta_{\mu\nu} =  ( H_{\mu\nu} + h_{\mu\nu})/m_p, \; \phi =  \Phi + \phi_{pot},  
\end{equation}
where $ H_{\mu\nu}$ and $\Phi$ are referred as the radiation modes, $h_{\mu\nu}$ and $\phi_{pot}$ as the potential modes.
In the following, we will use $\phi$ without subscription to refer to the potential part of the scalar field in the case of no confusion. 

In the gravitating binary system, our goal is to calculate the effective Lagrangian for the two stars and to give observable quantities. We consider the amplitude of two classically moving point particles scattering by exchanging potential modes. After integrating these potential modes, the amplitude will yield an effective Lagrangian of the classical motion of the 2 binary stars. Thus we can use the agreement of calculated quantities of the effective motion with the observation to determine a bound of our model parameters.

\subsection{Feynman rules of post-Newtonian EFT for binary system}

To perform the post-Newtonian EFT calculation, we need to derive Feynman rules from an expansion of the full Lagrangian and collect the relevant Feynman diagrams for a binary system. To calculate an effective Lagrangian of the compact stars up to 1PN order in an EFT approach, we derive the propagators of tensor and scalar modes, and relevant three-point vertices between these modes and worldline. Here we only show the propagators, and leave the vertices in the Appendix~\ref{sec:feyn}.
\begin{figure}
    \centering
    \includegraphics[width=\linewidth]{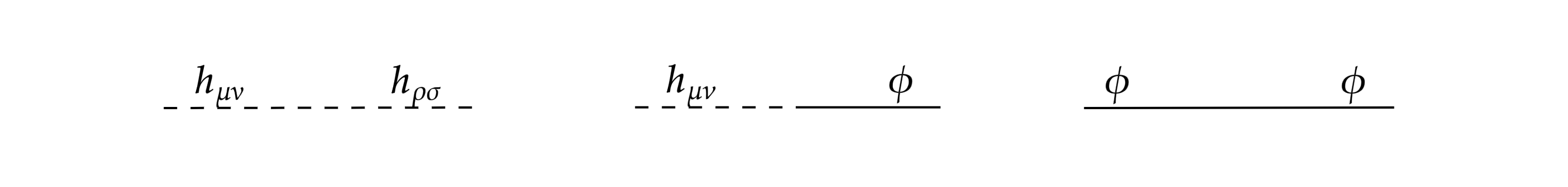}
    \caption{Tensor and scalar propagators. Solid line represents the scalar mode, and dashed line represent tensor mode.}
    \label{fig:prop}
\end{figure}

To obtain the propagators, we expand the action up to $\phi^2, \phi h$ and  $h^2$  order, and turn to momentum space representation,
 \begin{equation}
      h_{\mu\nu}(x) = \int_p e^{-i p \cdot x}h_{\mu\nu}(p),
 \end{equation} 
 where $\int_p$ stands for integration over momentum space$\int d^4 p/{(2\pi)^4}$. Then the relevant action can be written as follows,
\begin{equation}
\begin{aligned}
    i S_{\phi^2, \phi h, h^2} = \int_{p} \frac{i p^2}{2} \begin{pmatrix}
    h_{\mu\nu} & \phi \\    
    \end{pmatrix} 
    \begin{pmatrix}
        \mathbf{A}^{\mu\nu;\rho\sigma} &
        b^{\mu\nu} \\
        b^{\rho\sigma} &
        1 \\
    \end{pmatrix}
    \begin{pmatrix}
        h_{\rho\sigma} \\ \phi
    \end{pmatrix},
    \end{aligned}
\end{equation}
where we have defined the following quantities
\begin{equation}
    \mathbf{A}^{\mu\nu;\rho\sigma} = \frac12 \left( \eta^{\mu\rho} \eta^{\nu\sigma} + \eta^{\mu\sigma}\eta^{\nu\rho} - \eta^{\mu\nu} \eta^{\rho\sigma} \right), \quad b^{\mu\nu} = -2g_4 \left( \eta^{\mu\nu} -\frac{p^{\mu}p^{\nu}}{p^2} \right).  
\end{equation}

The inverse of the kinetic terms then gives the tensor and scalar propagators as in Fig.~\ref{fig:prop},
 \begin{equation}
    \begin{pmatrix}
       \left\langle h_{\mu\nu}(p) h_{\rho\sigma}(-p)\right\rangle &
       \left\langle h_{\mu\nu}(p) \phi(-p) \right\rangle \\
       \left\langle \phi(p) h_{\rho\sigma}(-p) \right\rangle & 
       \left\langle \phi(p) \phi(-p) \right\rangle
       \\
    \end{pmatrix} 
     = \frac{i}{p^2}\begin{pmatrix}
          \mathbf{A}^{-1}_{\mu\nu;\rho\sigma} + \frac{( \mathbf{A}^{-1}bb^{T} \mathbf{A}^{-1})_{\mu\nu;\rho\sigma}}{1-b^{T} \mathbf{A}^{-1}b} &
         - \frac{( \mathbf{A}^{-1} b)_{\mu\nu}}{1-b^{T}  \mathbf{A}^{-1} b} \\
         - \frac{(b^{T}  \mathbf{A}^{-1})_{\rho\sigma}}{1-b^{T}  \mathbf{A}^{-1} b}&
         \frac{1}{1-b^{T}  \mathbf{A}^{-1} b}
     \end{pmatrix},
 \end{equation} 
where we can calculate 
 \begin{align}
     &\mathbf{A}^{-1}_{\mu\nu;\rho\sigma} 
     = \frac12 \left( \eta^{\mu\rho} \eta^{\nu\sigma} + \eta^{\mu\sigma}\eta^{\nu\rho} - \eta^{\mu\nu} \eta^{\rho\sigma} \right),  
     q_{\mu\nu}(p)\equiv  \eta_{\mu\nu} + 2\frac{p_{\mu}p_{\nu}}{p^2}, \\
    & ( \mathbf{A}^{-1}b)_{\mu\nu} = g_4 q_{\mu\nu}(p), \; ( \mathbf{A}^{-1}bb^{T} \mathbf{A}^{-1})_{\mu\nu;\rho\sigma}  = g_4^{2} q_{\mu\nu}(p) q_{\rho\sigma}(p), \\
    &  b^{T}  \mathbf{A}^{-1} b  = -6g_4^{2}, \; 1-b^{T}  \mathbf{A}^{-1} b = 1+6g_4^2.
 \end{align}
Then we can read the propagator for the tensor modes,
\begin{align} 
    \left\langle h_{\mu\nu}(p) h_{\rho\sigma}(-p) \right\rangle 
    = \frac{i}{p^2}P_{\mu\nu;\rho\sigma}(p),\; P_{\mu\nu;\rho\sigma} = \mathbf{A}^{-1}_{\mu\nu;\rho\sigma}  -\frac{\kappa}{2} q_{\mu\nu}(p) q_{\rho\sigma}(p),\; \kappa  \equiv  \frac{2g_4^2}{1+6g_4^2}.
\end{align}
And the scalar-tensor mixing term is given by the off-diagonal term,
\begin{equation}
    \left\langle h_{\mu\nu}(p) \phi(-p) \right\rangle = -\frac{i }{p^2} \frac{g_4}{1+6g_4^2} q_{\mu\nu}(p) . 
\end{equation}
For off-shell potential modes, $p=(\omega, \mathbf{p}),\enskip \omega \sim |\mathbf{p}| v$. Hence in the leading order, we can estimate that
\begin{equation}
    \left\{  \begin{aligned}
    P_{00;00} =&\,\, \frac12 (1+\kappa), \\
    P_{00;ij} =&\,\, \frac12 (1-\kappa) \delta_{ij} - \kappa \frac{\mathbf{p}_i \mathbf{p}_j}{\mathbf{p}^2},\\
    P_{0i;0j} =&\,\, -\frac12 \delta_{ij} .
    \end{aligned} 
    \right.
\end{equation}
For the off-diagonal term, the components of $q_{\mu\nu}$ are 
\begin{equation}
    q_{00} = 1, \quad q_{0i} = 0, \quad q_{ij} = -\delta_{ij} - 2\frac{\mathbf{p_i}\mathbf{p_j}}{\mathbf{p}^2}
\end{equation}
at the leading order.
To perform the integration in potential modes, we turn to three-dimensional Fourier-transformed configuration space in which the propagators are written as
\begin{equation}
    \begin{pmatrix}
       \left\langle h_{\mu\nu}(\mathbf{p},t_1) h_{\rho\sigma}(-\mathbf{p},t_2)\right\rangle &
       \left\langle h_{\mu\nu}(\mathbf{p},t_1) \phi(-\mathbf{p},t_2) \right\rangle \\
       \left\langle \phi(\mathbf{p},t_1) h_{\rho\sigma}(-\mathbf{p},t_2) \right\rangle & 
       \left\langle \phi(\mathbf{p},t_1) \phi(-\mathbf{p},t_2) \right\rangle
       \\
    \end{pmatrix} 
     = -\frac{i}{\mathbf{p}^2}2\pi\delta(t_1-t_2)
     \begin{pmatrix}
         P_{\mu\nu;\rho\sigma}(\mathbf{p}) &
         -\dfrac{g_4q_{\mu\nu}(\mathbf{p})}{1+6 g_4^2}  \\
         -\dfrac{g_4q_{\rho\sigma}(\mathbf{p})}{1+6 g_4^2} &
         \dfrac{1}{1+6 g_4^2}
     \end{pmatrix}.
\end{equation}
Note that we may diagonalize this propagator matrix, which then induce coupling between matter and scalar field. However, in the basis without diagonalization the contributing diagrams would be simpler due to the absence of direct coupling between matter and scalar field, although the two bases are equivalent.

\begin{figure}
     \centering
     \includegraphics[width=\linewidth]{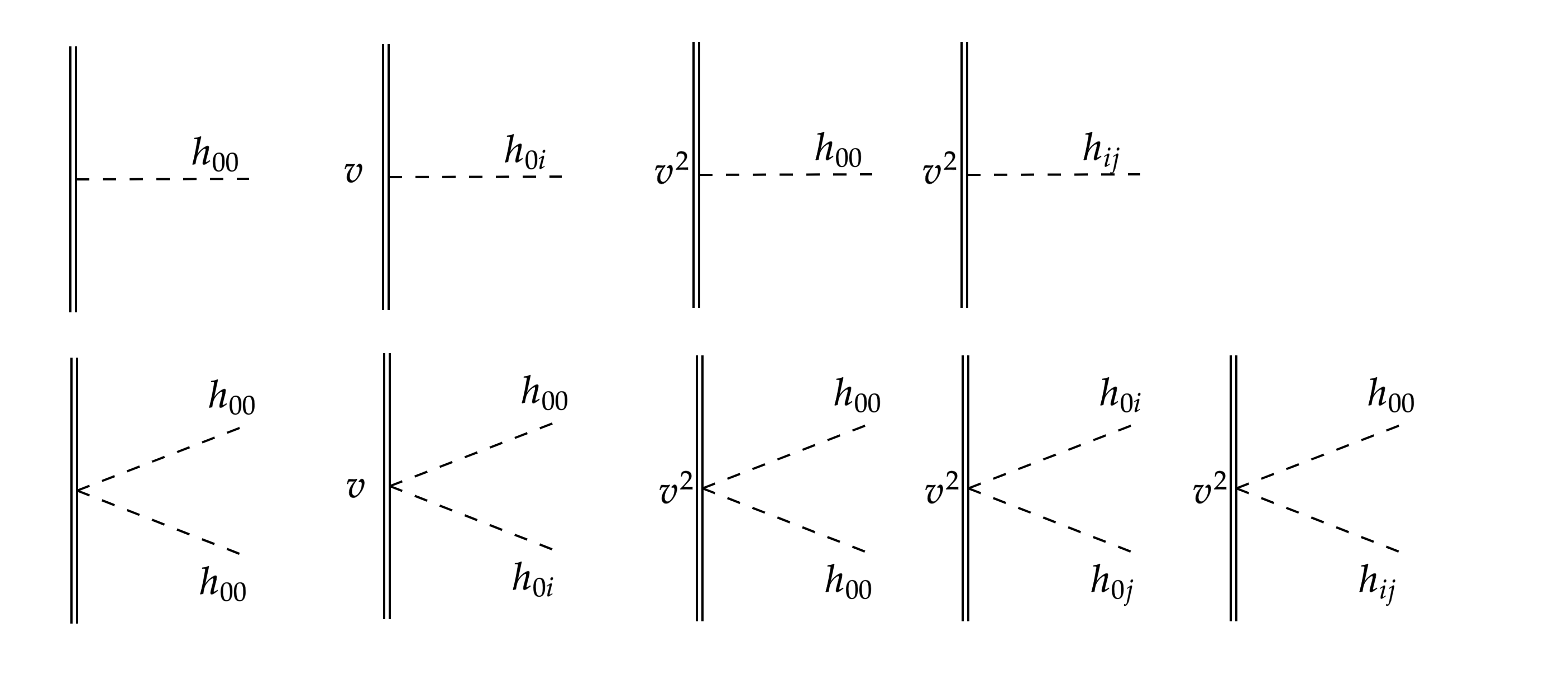}
     \caption{Worldline-graviton vertices. Double lines represent particle worldline while dashed lines represent (tensor mode) gravitons.}
     \label{fig:WL-Vertices}
 \end{figure}

\subsection{Extraction of the classical contribution}

As we need the classical dynamics of in the end, we have to extract the pure classical part (leading in $\hbar$) from the EFT calculations. In other words, this means that we should limit the exchanged graviton modes to the potential modes. Such contributions involve the diagrams satisfying 2 criteria:
\begin{enumerate}
    \item Every loop must include an edge on one of the worldlines, otherwise the integration of loop momenta at a large energy scale will introduce radiative contribution.
    \item No internal edge starting and ending on the same worldline, otherwise there will be internal energy contribution for the particle worldline.
\end{enumerate}
With this restriction, the number of relevant diagrams can be considerably reduced. 

When integrating internal momenta in loop diagrams, we have to regularize and renormalize the divergences in the divergent integrals. We choose dimensional regularization and minimal subtraction scheme in EFT approach~\cite{Goldberger:2004jt}. So the dimensionless integrals vanishes and can be dropped. For 1PN calculations, the most general integrals can be written as following,
\begin{equation}
    f(n_1,n_2,n_3,n_4,n_5;r) =\int \frac{\mathrm{d}^d p_1}{(2\pi)^d} \frac{\mathrm{d}^d p_2}{(2\pi)^d} \frac{ (p_1 \cdot r)^{-n_4}(p_2 \cdot r)^{-n_5}}{\left(p_1^2\right)^{n_1} \left(p_2^2\right)^{n_2} \left[(p_1+p_2)^2\right]^{n_3}} e^{i p_1 \cdot r + i p_2 \cdot r}.
\end{equation} 
We set the dimension $d=3$ and drop the scaleless integrals. Among integrals with different $n_i$, there are linear relations generated from the integration of full derivatives or integration by part. Using these relations, all the integrals are reduced to $f(1,1,0,0,0;r)$ multiplied by a factor. After integration, $f(1,1,0,0,0;r)$ yields simply
\begin{equation}
    f(1,1,0,0,0;r) = \int \frac{\mathrm{d}^d p_1}{(2\pi)^d} \frac{\mathrm{d}^d p_2}{(2\pi)^d} \frac{e^{i p_1 \cdot r}}{p_1^2} \frac{e^{i p_2 \cdot r}}{p_2^2} =\left(\frac{1}{4\pi r}\right)^2, \quad d \to 3.
\end{equation}
For example, we have
\begin{align}
    f(1,1,-1,0,0;r) &= -\frac{2}{r^2}f(1,1,0,0,0;r) = -\frac{1}{8\pi^2} \frac{1}{r^4}, \\
    f(1,1,-2,0,0;r) &= \frac{24}{r^4}f(1,1,0,0,0;r) = \frac{3}{2\pi^2} \frac{1}{r^6}.
\end{align} 

\section{Binary dynamics in Horndeski theory}\label{sec:PN}
In this section, we discuss the classical binary dynamics in Horndeski gravity up to 1PN and show the leading contributions in diagrammatic pictures. By calculating the scattering amplitude, we can obtain the corresponding Lagrangian and Hamiltonian, which determine the classical dynamics. 

\subsection{Post-Newtonian Results}
\begin{figure}
    \centering
    \includegraphics[width=\linewidth]{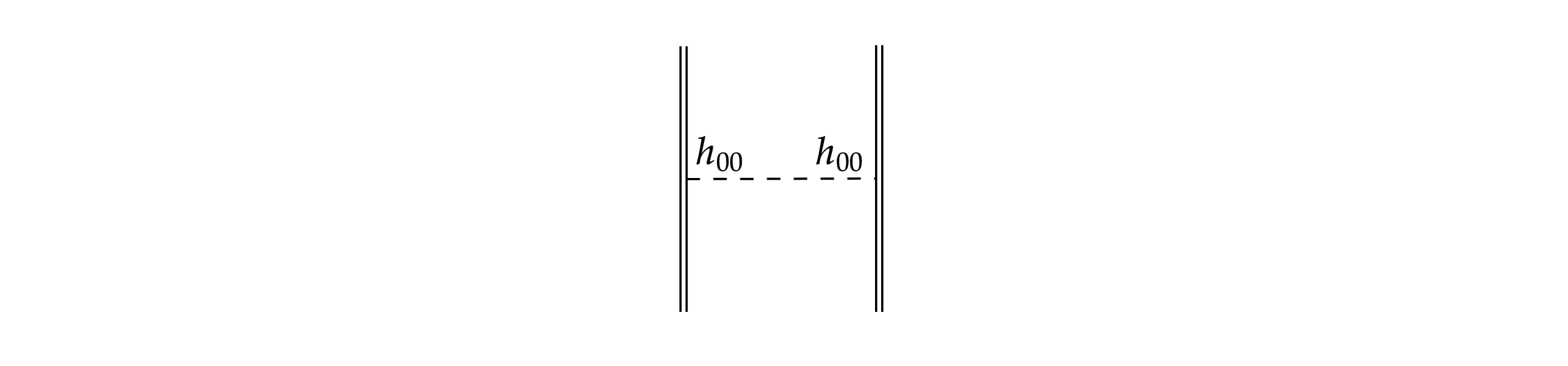}
    \caption{The leading-order diagram for binary dynamics.}
    \label{fig:LO}
\end{figure}
\begin{figure}
    \centering
    \includegraphics[width=\linewidth]{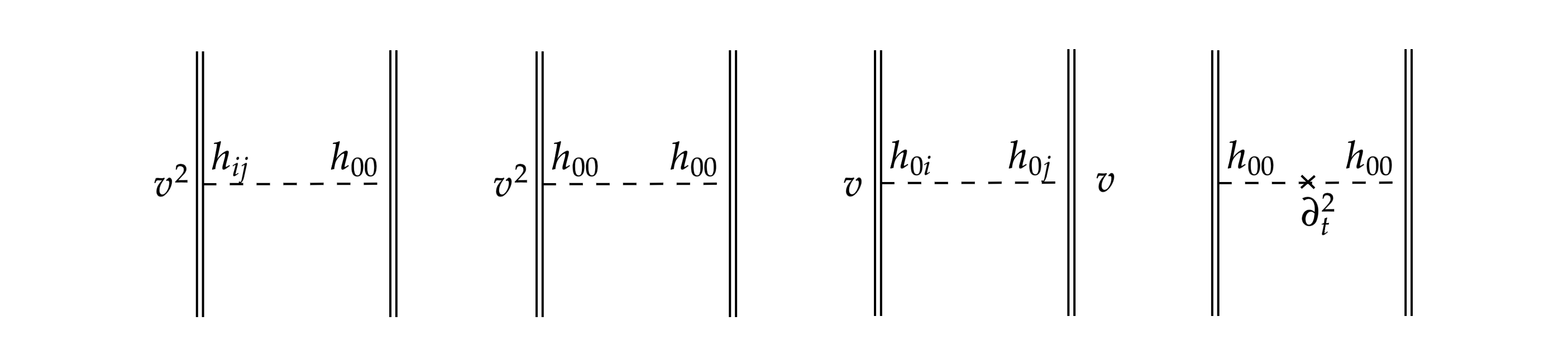}
    \caption{Next-to-leading diagrams at tree level with tensor modes only. (a)}
    \label{fig:NLO-a}
\end{figure}

\begin{figure}
    \centering
    \includegraphics[width=\linewidth]{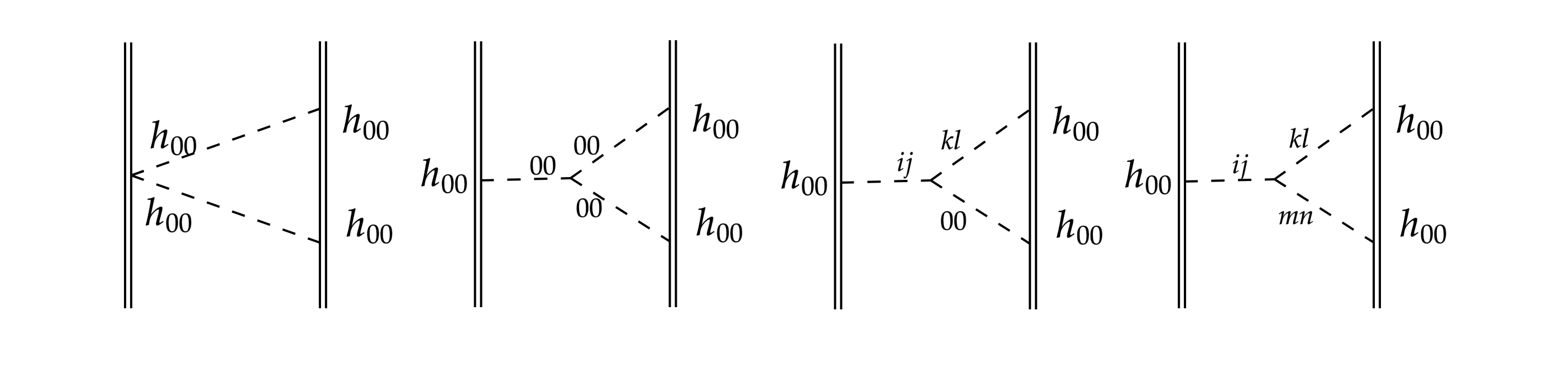}
    \caption{Next-to-leading diagrams at one-loop level with tensor modes only. (b) }
    \label{fig:NLO-b}
\end{figure}

At the Newtonian order, we have only one diagram in Fig.~\ref{fig:LO} for classical binary dynamics. The amplitude of Fig.~\ref{fig:LO} is calculated as
\begin{equation}
    i\mathcal{A}_{0}  = i\int \mathrm{d}t \,\frac{Gm_1m_2}{r}(1+\kappa).
\end{equation} 
At the first post-Newtonian (1PN) order, we have the following diagrams to calculate. The diagrams in Fig.~\ref{fig:NLO-a} have the same topology as the leading order but with higher order $O(v^2)$ vertices, thus they scale as $O(Gv^2)$. We calculate the amplitude as
\begin{equation}
\begin{aligned}
    i\mathcal{A}_{1a} 
    =& i\int \mathrm{d}t \, \frac{G m_1 m_2}{r} \left[ \frac32 \left(1+\frac{\kappa}{3}\right)\left(\mathbf{v}_1^2+\mathbf{v}_2^2\right)
-\frac72 \left( 1+\frac{3\kappa}{7}\right) \mathbf{v}_1 \cdot \mathbf{v}_2 \right. \\
& \left. - \frac12 (1-3\kappa) (\mathbf{n}\cdot \mathbf{v}_1) (\mathbf{n} \cdot \mathbf{v}_2)
+\frac{\kappa}{2} \left( (\mathbf{n}\cdot \mathbf{v}_1)^2 + (\mathbf{n} \cdot \mathbf{v}_2)^2 \right)
\right], \mathbf{n}\equiv \mathbf{r}/r .
\end{aligned}
\end{equation}
The diagrams in Fig.~\ref{fig:NLO-b} are one-loop ones with leading-order vertices and scale as $O(G^2)$. We obtain
\begin{equation}
    i\mathcal{A}_{1b} 
    = i\int \mathrm{d}t \, \frac{G^2 m_1 m_2(m_1+m_2)}{r^2} \left(-\frac12 \right)(1 - 7\kappa + 5\kappa^2 - 5 \kappa^3). 
\end{equation}

\begin{figure}[t]
    \centering
    \includegraphics[width=\linewidth]{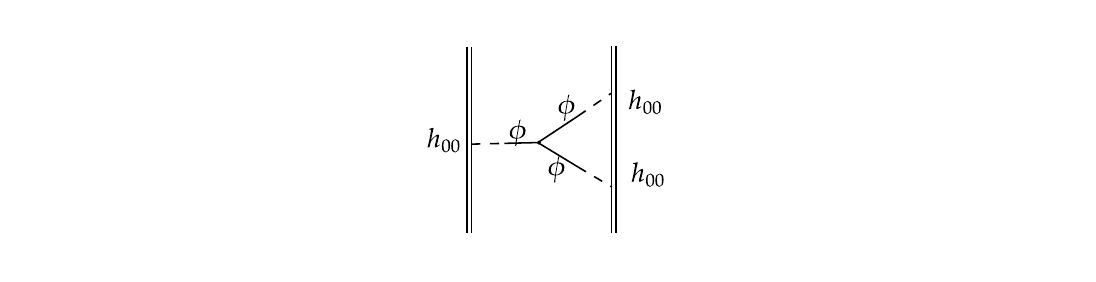}
    \caption{The next-to-leading diagram at one-loop level with scalar self-interaction. (c) }
    \label{fig:NLO-c}
\end{figure}

\begin{figure}
    \centering
    \includegraphics[width=\linewidth]{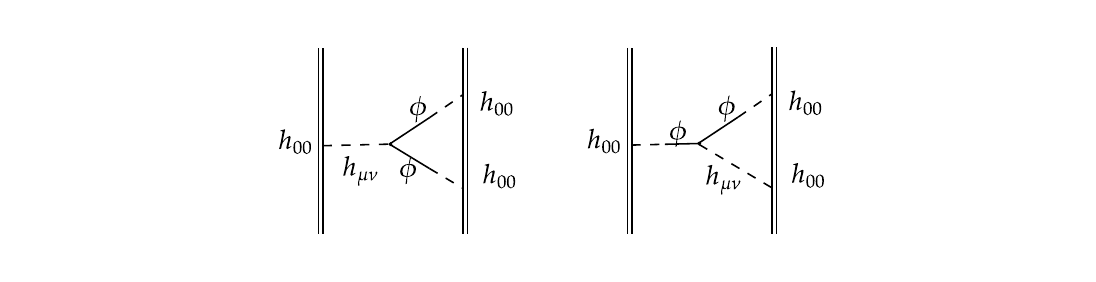}
     \caption{Next-to-leading diagrams at one-loop level with graviton-scalar mixing. (d)}
        \label{fig:NLO-d}
\end{figure}
\begin{figure}
	\centering
    \includegraphics[width=\linewidth]{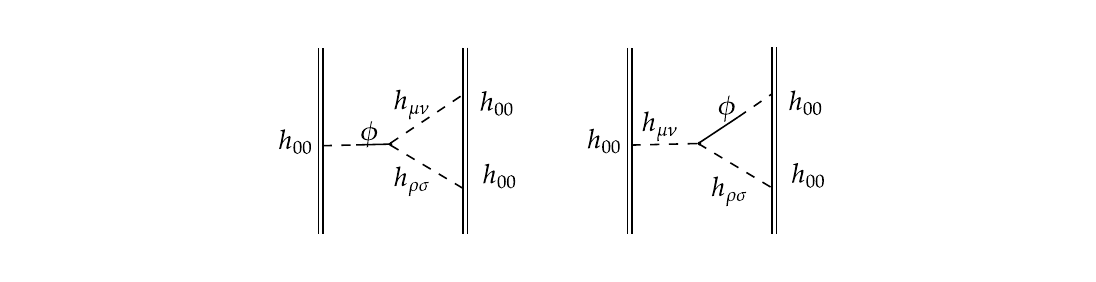}
    \caption{Next-to-leading diagrams at one-loop level with graviton-scalar mixing. (e)}
    \label{fig:NLO-e}
\end{figure}

Because of the $\phi^3$ vertex (spatial derivatives are allowed), we also have diagram Fig.~\ref{fig:NLO-c}. As we are working in the base with $\phi-h$ mixing-type propagators, only $h$ gravitons are interacting with matter worldline directly. Therefore, this diagram is not vanishing when the mixing is present. We can obtain
\begin{equation}
    i\mathcal{A}_{1c} 
    = i\int \mathrm{d}t \, \frac{G^2 m_1 m_2(m_1+m_2)}{r^2} \left( -4 g_3\frac{l_{p}^2}{r^2} \right)\left( \frac{g_4}{1+6g_4}\right)^3,\;\;
    l_{p} = \sqrt{\frac{G\hbar}{c^3}}.
\end{equation}
This contribution is proportional to $1/r^4$ and comes from the scalar self-coupling with derivatives $g_3 X\square\phi$. 
Note that the presence of Planck scale $l_{p}^2$ does not mean it is a quantum correction, but because our parameterization of $g_3$ in Eq.~\ref{eq:parameters} introduces such a length scale for such a coupling. Moreover, the coupling $g_2\phi^3$ does not contribute at this leading order, since it is proportional to the integral
\begin{equation}
    \int 
    \frac{\mathrm{d}^3\mathbf{p}_1}{(2\pi)^3} \frac{\mathrm{d}^3\mathbf{p}_2}{(2\pi)^3} \frac{1}{\mathbf{p}_1^2 \mathbf{p}_2^2 (\mathbf{p}_1+\mathbf{p}_2)^2}
    e^{i\mathbf{p}_1\cdot r+i\mathbf{p}_2\cdot r}.
\end{equation} 
This is scaling as $p^0$ and hence the result is a constant in $r$ and does not affect the dynamics.

Additionaly, there are diagrams in Fig.~\ref{fig:NLO-d} and Fig.~\ref{fig:NLO-e} contributing. Their results are respectively
\begin{equation}
    i\mathcal{A}_{1d} 
    = i\int \mathrm{d}t \, \frac{G^2 m_1 m_2(m_1+m_2)}{r^2} \frac{\kappa}{2}\left( 1-3\kappa \right),
\end{equation}
and
\begin{equation}
    i\mathcal{A}_{1e} 
    = i\int \mathrm{d}t \, \frac{G^2 m_1 m_2(m_1+m_2)}{r^2} \kappa\left( 1 -6 \kappa + \frac72 \kappa^2  \right).
\end{equation}

\subsection{Reduced Hamiltonian and Periastron Advance}
In the amplitudes we have calculated, we should make the substitution $ G(1+\kappa) \to G $, so that the Newtonian-order effective lagrangian is the same as Einstein's effective lagrangian for two particles. The effective Lagrangian is extracted from the matching relation
\begin{equation}
   i \mathcal{A} = i \int L \, \mathrm{d}t .
\end{equation}
Exploiting the matching relation at each PN order, we have
\begin{align} 
   & L_{0PN} = \sum_a \frac12 m_a \mathbf{v}_a^2 + \frac{G m_1 m_2}{r}, \\
   & L_{1PN} = \sum_a \frac18 m_a \mathbf{v}_a^4 + \frac{G^2m_1 m_2 (m_1+m_2)}{r^2} \left(\alpha + \tilde{\alpha}\frac{l_{p}^2}{r^2} \right)  \nonumber \\
    &  + \frac{G m_1 m_2}{r} \left\{ \frac{}{} \beta (\mathbf{v}_1^2 + \mathbf{v}_2^2) + \gamma \mathbf{v}_1 \cdot \mathbf{v}_2 + \delta (\mathbf{n}\cdot \mathbf{v}_1)(\mathbf{n} \cdot \mathbf{v}_2) + \epsilon \left[ (\mathbf{n}\cdot \mathbf{v}_1)^2 + (\mathbf{n} \cdot \mathbf{v}_2)^2 \right] \right\},
\end{align}
where we have defined the following quantities,
\begin{align}
    \alpha &= -\frac{1 - 10 \kappa + 20 \kappa^2 -12 \kappa^3}{2(1+\kappa)^{2}}, \;
    \tilde{\alpha} = -4 g_3 \left( \frac{g_4}{1+6g_4^2} \right)^3/ (1+\kappa)^{2},\; 
    \beta = \left( \frac32 + \frac12 \kappa \right)/(1+\kappa),  \; \nonumber \\
    \gamma &= \left( -\frac72 - \frac32 \kappa\right)/(1+\kappa), \;
    \delta = \left( -\frac12 + \frac32 \kappa\right)/(1+\kappa), \;
    \epsilon = \frac{\kappa}{2(1+\kappa)}.
\end{align}

After a Legendre transformation of the 1PN two-body effective Lagrangian, we can get the Hamiltonian. In the meantime, the two-body motion can be reduced to the motion of a reduced mass $ \mu = m_1m_2/(m_1+m_2)$. After rescaling the following quantities,
\begin{equation}
    p \to p/\mu, \quad r \to r/(GM), \quad E \to E/\mu, \quad t \to t/(GM), 
\end{equation}
we have the Hamiltonian for the motion of reduced mass
\begin{equation}
\begin{aligned}
         H =&\,\, \frac{\mathbf{p}^2}{2} - \frac{1}{r} + \frac{1}{c^2} \left\{ -\frac18 (1-3\nu) \mathbf{p}^4  
         - \frac{1}{r^2} \left( \alpha + \tilde{\alpha} \frac{l_{p}^2}{r^2} \right) 
         \right. \\
         & \left. -\frac{1}{r} \left[ \left(\beta - (2\beta+ \gamma) \nu \right) \mathbf{p}^2 
         + \left( \epsilon - (2\epsilon + \delta) \nu \right) (\mathbf{p}\cdot \mathbf{n})^2 \right] \right\},
\end{aligned}
\end{equation}
where
\[ \mu = \frac{m_1 m_2}{m_1 + m_2}, \quad M = m_1 + m_2, \quad \nu = \frac{\mu}{M}. \]

According to Damour \& Schafer \cite{Damour:1988mr}, we can use Hamilton-Jacobi Formalism to derive the periastron advance $\Delta \Phi$ per period from the reduced Hamiltonian, $k \equiv {\Delta \Phi}/({2\pi})$,
\begin{equation}
    k =\frac{1}{c^2} 
    \left\{ \left[ \left( \frac12 + \alpha + 2\beta + \epsilon \right) - \left(\frac32 + 4\beta +2 \gamma + \delta + 2\epsilon \right)\nu \right]\frac{1}{h^2} + \frac{\tilde{\alpha}}{(GM)^2} \left[ \frac{3E}{h^4} + \frac{15}{2} \frac{1}{h^6} \right] \right\},
\end{equation}
where 
\[  h = \frac{J}{\mu G M} = \sqrt{\frac{(1-e^2)a}{GM}}, \quad E = -\frac{1}{\mu}\frac{GM\mu}{2a} = - \frac{GM}{2a}. \]
Rewriting with the observable parameters, eccentricity $e$, orbital period $P_b$ and the inferred total mass $M$, with the following relations,
\begin{equation}
    \left\{ \begin{aligned}
h^{-2}c^{-2} &= \frac{1}{1-e^2} \frac{GM}{a} = \frac{1}{1-e^2}(GM)^{\frac23} \left( \frac{P_b}{2\pi} \right)^{-\frac23}, \\
|E|c^{-2} &= \frac12 \frac{GM}{a} = \frac12 (GM)^{\frac23} \left(\frac{P_b}{2\pi}\right)^{-\frac23},
\end{aligned} \right.
\end{equation} 
we obtain the periastron advance in the discussed Horndeski theory,
\begin{equation}\label{eq:peri}
     k = \frac{1}{c^2 h^2} \left[\frac{3 + 21\kappa/2 - 8 \kappa^2 + 6 \kappa^3}{(1+\kappa)^{2}} - \frac{3 \kappa \nu}{1+\kappa}  \right]
    +  \frac{\tilde{\alpha} l_{p}^2 c^4}{(GM)^2} 
    \frac{6(1+e^2/4)}{(1-e^2)^3} \left( \frac{2\pi GM}{P_b} \right)^{2}.
\end{equation} 

%\section{Comparing with Experiment Results}

\section{Constraints on model parameters from Observation}\label{sec:numeric}
\begin{table}[t]
    \centering
    \begin{tabular}{c|c|c}\hline
        Quantity       & PSR B 1534+12 & PSR J0737-3039A/B \\ \hline
        Period & $P_b = 0.420737299122(10) \,d$ & $P_b = 0.1022515592973(10) \,d$ \\
        Eccentricity & $e = 0.2736775(3)$ & $e = 0.087777023(61) $ \\
        Pulsar Mass & $m_1 = 1.3332(10) \,M_{\odot}$ & $m_A = 1.338185(^{+12}_{-14}) \,M_{\odot}$ \\
        Companion Mass & $m_2 = 1.3452(10) \,M_{\odot}$ & $m_B = 1.248868(^{+13}_{-11}) \,M_{\odot}$ \\
        Total Mass & $M = 2.678428(18) \,M_{\odot}$  & $M = 2.587052(^{+9}_{-7}) \,M_{\odot} $  \\
        Precession Rate & $\dot{\omega} = 1.755789(9) \,deg/yr$ & $\dot{\omega} = 16.899323(13) \, deg/yr $ \\ \hline
    \end{tabular}
    \caption{Observables fitted from pulsar timing data of PSR B 1534+12~\cite{Fonseca:2014qla} and PSR J0737-3039~\cite{Lyne:2004cj}. }
    \label{tab:PSRdata}
\end{table}

Now we are in a position to use the observation data from several binary systems to constrain the model parameters. Here, we choose two typical pulsar binary systems, PSR B 1534+12~\cite{Fonseca:2014qla} and PSR J0737-3039 A/B~\cite{Lyne:2004cj}, which have a rather high observation precision of periastron advance. The observed data of the two systems are shown in the Tab.~\ref{tab:PSRdata}. We use the observed values of periastron advance and its error to obtain a bound for physical values in the $g_3-g_4$ plane. In principle, the total radiation power is another observable that can test the 1PN dynamics of the model. However, the precision is much lower than the periastron advance, therefore here we only use information from periastron advance.

From the observed pulsar timing data of the system, people usually extract a set of Kepler parameters from the data 
\begin{equation}
\{ \omega, x, e, T_0, P_b \},
\end{equation}
to represent the dynamics at the leading order of relativistic expansion \cite{Damour:1991rd}. Here, $\omega$ is the angle between the line from the orbital center to the point of periastron and the nodal line,  $x=a\sin i$ is the apparent semi-axis, and $T_0$ is the observation time. These experimental parameters can determine a set of classical conserved quantities $E, \mathbf{L}$ and the initial time $t_0$. The higher-order dynamics which is relevant in our model can be expressed as the rate of these experimental parameters. This so-called post-Keplerian parameters include the periastron advance $\dot{\omega}_0$ as well as the changing rate of the orbital period $\dot{T}_0$, which is related to the radiated power and other parameters. we will make use of the observation data of $\dot{\omega}_0$ to constrain our model parameters.

\begin{figure}[t]
        \centering
        \subfigure[\enskip PSR B 1534+12]{
        \begin{minipage}{0.48\textwidth}
        \centering
        \includegraphics[width=\textwidth]{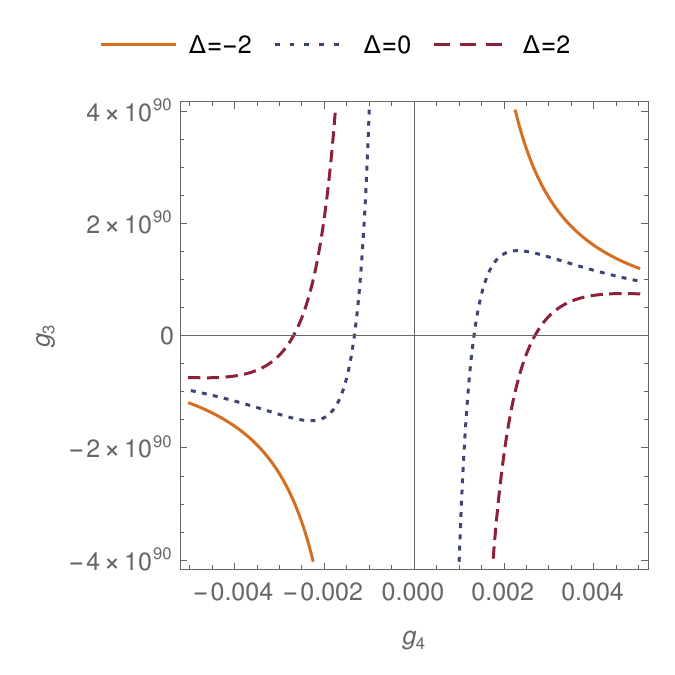} %\caption{PSR B 1534+12}
        \label{fig:PSRB1534+12}
        \end{minipage}   }
        \subfigure[\enskip PSR J0737-3039 A/B]{
        \begin{minipage}{0.48\textwidth}
        \centering
        \includegraphics[width=\textwidth]{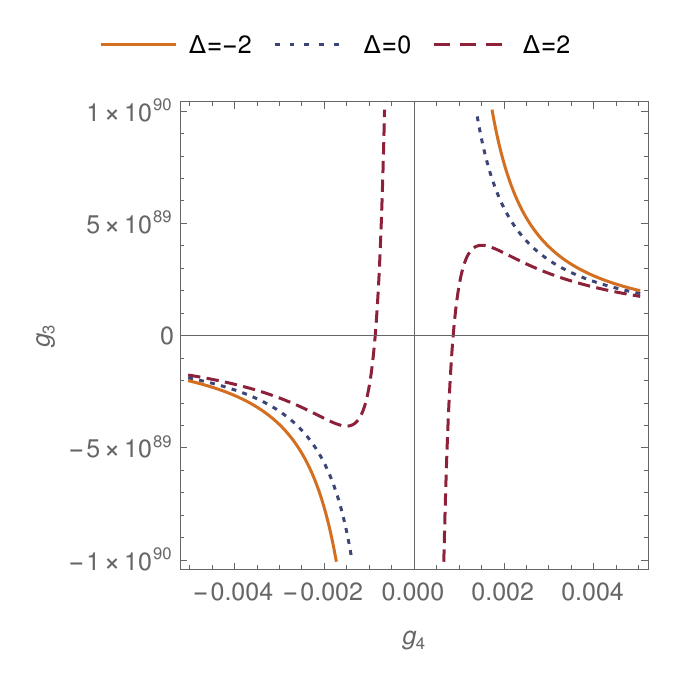}    %\caption{PSR J0737-3039 A/B}
        \label{fig:PSRJ0737-3039}
         \end{minipage}}
        \caption{Constraint of $g_3$ and $g_4$ from the two systems. The solid lines stand for the trajectory of $(g_4,g_3)$ values which give $k_{model}=k_{obs}-2\Sigma$, dotted lines stand for the trajectory of $k_{model}=k_{obs}$, and the dashed lines stand for the trajectory of $k_{model}=k_{obs}+2\Sigma$. \label{fig:constraint}}
\end{figure}

We implement the constraints from observation for fractional periastron advance per period $k$: 
\begin{equation}
    k_{obs} = \dot{\omega}\frac{P_b}{2\pi}, \;
    \Delta \equiv \frac{k_{model} - k_{obs}}{\sqrt{\sigma_{model}^2 + \sigma_{obs}^2}}, \; \left| \Delta \right| < 2, 
\end{equation} 
where $k_{model} \text{ is given by Eq.~\ref{eq:peri}},$ and should lie within $2\sigma$ around the observed value $k_{obs}$.

The bound of $g_3$ and $g_4$ is showed in Fig.~\ref{fig:PSRB1534+12} and Fig.~\ref{fig:PSRJ0737-3039} from the two systems, respectively. We use $\Sigma = \sqrt{\sigma_{model}^2 + \sigma_{obs}^2}$ here to represent the total standard error. As the observed values can be greater or smaller than the center value of predicted ones, we see the dotted lines have different shapes in these two systems. The reason is that in Fig.~\ref{fig:PSRB1534+12} we have $k_{model,GR} < k_{obs}$, while in Fig.~\ref{fig:PSRJ0737-3039} we have $k_{model,GR} > k_{obs}$. Here $k_{model,GR}$ is the predicted observables if there is no corrections for general relativity. Despite of this difference, the predicted and observed values in the two examples both lie within the $2\sigma$ range. Note that the large value of $g_3$ is again due to our parametrization of $g_3$ terms in Eq.~\ref{eq:parameters}.

\section{Conclusion}\label{sec:discussion}
We have investgitated the conservative dynamics of a binary system up to the first-order post-Newtonian in Horndeski gravity by formulating its effective field theory. From the effective Lagrangian or Hamiltonian, we have computed the periastron advance of a binary system in its inspiral phase. Comparing the theoretical predictions of periastron advance for two binary systems with their obeservations, PSR B 1534+12 and PSR J0737-3039, we have obtained the constraints on the model parameters, $g_3$ and $g_4$ in Eq.~\ref{eq:parameters}. The contours of the available parameter space are shown in Fig.~\ref{fig:constraint}. As the precision observation will improve in the future, we expect the bound of these parameters will be more restrictive. Also, the formalism developed here can also been extended to higher orders systematically by summing higher-loop diagrams and applied to other binary systems, including binary black holes. Effects on the gravitational wave of such inspiral system would also be interesting~\cite{Bernard:2018hta, Bernard:2018ivi, Bernard:2022noq, Shiralilou:2021mfl, Julie:2022qux}, which we shall explore for future work. 

\section*{Acknowledgement}
This work is supported by National Key Research and Development Program of China (Grant No.2021YFC2201901), and National Natural Science Foundation of China under Grants No.12147103 and 11851302.

\appendix
\section{$\phi X$ term and redefinition}\label{sec:app1}
The $g_2 \phi X$ term can be removed by field redefinition $\phi \to \Phi$ with 
\begin{equation}
    (1+ \phi) X_{\phi} = X_{\Phi}.
\end{equation}
Here we use the notation $X_A=\frac12 g^{\mu\nu}\nabla_{\mu} A \nabla_{\nu} A$ and unit $m_p = 1$. In differential form, we demand that
\begin{equation}
    \sqrt{1+g_2 \phi}\,d\phi \to d\Phi. 
\end{equation}
To integrate the equation, we set the zero point of $\Phi$ so that $\Phi=0$ when $\phi=0$, then
\begin{align}
    \phi &= g_2^{-1} \left[ \left( 1 + \frac32 g_2 \Phi \right)^{\frac23}-1\right], \\
    X_{\phi} &= \left(1+\frac32 g_2 \Phi\right)^{-\frac23}X_{\Phi}, \\
    \square \phi &= \left(1+\frac32 g_2 \Phi\right)^{-\frac13} \square \Phi - g_2 \left(1+\frac32 g_2 \Phi\right)^{-\frac43} X_{\Phi}, \\
    X_{\phi} \square \phi &= \left(1+\frac32 g_2 \Phi\right)^{-1} X_{\Phi} \square \Phi 
    - g_2 \left(1+\frac32 g_2 \Phi \right)^{-2} X_{\Phi}^2.
\end{align}
Assuming that this correction to kinematic term is small $g_2 \phi \ll 1$, 
\begin{equation}
    \phi = \Phi - \frac14 g_2 \Phi^2 + o(\Phi^2).
\end{equation}
Then the correction to $G_3$ is of higher order,
\begin{equation}
        X_{\phi} \square \phi 
        = \left(1 - \frac32 g_2 \Phi + o(\Phi)\right) X_{\Phi} \square \Phi 
        - g_2 \left(1 - 3 g_2 \Phi + o(\Phi) \right) X_{\Phi}^2.
\end{equation}

\section{Feynman Rules} \label{sec:feyn}
Similarly to the derivation of graviton propagators, we expand the worldline action $S_{pp}$ to find the couplings of gravitons and the point mass source at each order. To calculate an effective Lagrangian up to 1PN order, we need worldline vertices with at most two gravitons and up to $O(v^2)$ order. We expand the Lagrangian to get 
\begin{equation}
    \begin{aligned} \label{eq:Spp}
    S_{pp} = &- \sum_a \int m_a d\tau_a = - \sum_a \int m_a \sqrt{g_{\mu\nu}\frac{dx^{\mu}}{dt} \frac{dx^{\nu}}{dt}} dt \\
      =& - \sum_a \int dt\, m_a \biggl( 
    1 + \frac{h_{00}}{2m_p} 
    - \frac{h_{00}^2}{8m_p^2} 
    + \frac{h_{0i}v_a^i}{m_p} 
    - \frac{h_{00}h_{0i}v_a^i}{2m_p^2} \\
    & - \frac{v_a^2}{2} + \frac{h_{00} v_a^2}{4m_p} + \frac{h_{ij}v_a^i v_a^j}{2m_p} 
    - \frac{3 h_{00}^2 v_a^2}{16 m_p^2} 
    - \frac{h_{00} h_{ij} v_a^i v_a^j}{4 m_p^2} 
    - \frac{h_{0i} h_{0j} v_a^i v_a^j}{2 m_p^2}
    + \cdots \biggr).
    \end{aligned}
\end{equation}
The Lagrangian of particle worldlines in Eq.~\ref{eq:Spp} are expanded up to $O(v^2)$ terms. Terms without $h_{\mu\nu}$ contribute to the kinetic energy of the binary. At 1PN order, this yields the kinetic energy
\begin{equation}
    E_{kin} = \sum_{a=1,2} \frac12 m_a \mathbf{v}_a^2 + \sum_{a=1,2} \frac18 m_a \mathbf{v}_a^4.
\end{equation}
Up to $v^2 h^2$ order, we have the worldline-graviton vertices as in Fig.~\ref{fig:WL-Vertices}, these vertices are summarized in the Tab.~\ref{tab:worldline}.
\begin{table}
    \centering
    \[\begin{array}{r|c|c|c}
 \hline & O(v^0) & O(v^1) & O(v^2)\\
 \hline
    h^1 & \displaystyle h_{00}:\enskip 
    -i\frac{m_a}{2m_p} e^{i \mathbf{k}\cdot \mathbf{x}_a(t)}  &
    \displaystyle h_{0i}:\enskip 
    -i\frac{m_a v_a^i}{m_p} e^{i \mathbf{k}\cdot \mathbf{x}_a(t)} &
    \parbox[c]{12em}{
  \begin{equation*}
    \begin{matrix}
	\displaystyle h_{00}: \enskip 
	-i\frac{m_a v_a^2}{4m_p}e^{i \mathbf{k}\cdot \mathbf{x}_a(t)} \\
	\displaystyle h_{ij}: \enskip 
	-i\frac{m_a v_a^i v_a^j}{2m_p} e^{i \mathbf{k}\cdot \mathbf{x}_a(t)}
    \end{matrix}
  \end{equation*}
} 
\\ \hline
    h^2 & \displaystyle h_{00}^2:\enskip 
    i\frac{m_a}{4m_p^2}e^{i(\mathbf{k_1}+\mathbf{k}_2)\cdot \mathbf{x}_a(t)} &
    \displaystyle h_{00}h_{0i}:\enskip
     i\frac{m_a v_a^i}{2m_p^2} e^{i (\mathbf{k_1} + \mathbf{k_2})\cdot \mathbf{x}_a(t)} &
     \parbox[c]{12em}{
  \begin{equation*}
     \begin{matrix}
	\displaystyle h_{00}^2:\enskip 
	i\frac{3m_a v_a^2}{8m_p^2} 
	e^{i (\mathbf{k_1} + \mathbf{k_2})\cdot \mathbf{x}_a(t)} \\
	\displaystyle h_{0i}h_{0j}:\enskip 
	i\frac{m_a v_a^i v_a^j}{m_p^2} e^{i (\mathbf{k_1} + \mathbf{k_2})\cdot \mathbf{x}_a(t)}  \\
	\displaystyle h_{00}h_{ij}:\enskip 
	i\frac{m_a v_a^i v_a^j}{4m_p^2}  e^{i (\mathbf{k_1} + \mathbf{k_2})\cdot \mathbf{x}_a(t)}     
    \end{matrix}
  \end{equation*}
} \\ \hline
 \end{array} \]    
 \caption{Worldline-graviton couplings up to $O(h^2)$, $O(v^2)$.}
    \label{tab:worldline}
\end{table}

For the three-point gravition self-interaction vertices, we need to expand the action to $O(h^3)$ order. The action is then Fourier transformed and symmetrized in $p_1,p_2, p_3$:
\begin{align}
    i S_{h^3} &= -i\frac{2}{m_p} \frac{1}{3!} \int_{p_i} \delta_{p_i} \left\{ \frac14 \Bigl[ \
h(p_1)h(p_2)h_{\mu\nu}(p_3)p_1^{\mu} p_2^{\nu} + (1\rightarrow2\rightarrow3\rightarrow1) + (1\rightarrow3\rightarrow2\rightarrow1) \Bigr] 
\right.\nonumber\\
& +\left[\frac{1}{16} h(p_1) h(p_2) h(p_3)-\frac18 h(p_1) h^{\mu\nu}(p_2) h_{\mu\nu}(p_3)+ \frac12 h_{\mu\nu}(p_1) h^{\mu}_{\enskip\rho}(p_2) h^{\nu}_{\enskip\sigma} (p_3) \right]\sum_i p_i^2 \nonumber\\
& + \frac12 \Bigl[ 
h(p_1) h_{\mu}^{\enskip\rho}(p_2) h_{\nu\rho}(p_3) (p_2^{\mu} p_3^{\nu} - p_3^{\mu} p_2^{\nu}) 
 + (1\rightarrow2\rightarrow3\rightarrow1) + (1\rightarrow3\rightarrow2\rightarrow1) \Bigr] \nonumber\\
&- \frac12 \Bigl[ 
h_{\mu\nu}(p_1) h^{\rho\sigma}(p_2) h_{\rho\sigma}(p_3) p_2^{\mu} p_3^{\nu} 
+  (1\rightarrow2\rightarrow3\rightarrow1) + (1\rightarrow3\rightarrow2\rightarrow1) \Bigr]\nonumber \\
& + \left. \frac12 \Bigl[ 
h_{\rho\sigma}(p_1) h_{\mu}^{\enskip\rho}(p_2) h_{\nu}^{\enskip\sigma}(p_3) (p_2^{\mu} p_3^{\nu} - p_3^{\mu} p_2^{\nu})
+ (1\rightarrow2\rightarrow3\rightarrow1) + (1\rightarrow3\rightarrow2\rightarrow1) \Bigr]  \right\},
\end{align}
where $h$ without indices stands for traces $h(p)=h^{\enskip\mu}_{\mu}(p)$ and
\[ \delta_{p_i} = (2\pi)^4\delta(p_1+p_2+p_3), \quad \int_{p_i} (\cdots) = \int \frac{\mathrm{d}^4 p_1}{(2\pi)^4} \frac{\mathrm{d}^4 p_2}{(2\pi)^4} \frac{\mathrm{d}^4 p_3}{(2\pi)^4} (\cdots). \]
Similarly, we have symmetrized 3-point vertices in the expanded action at $O(\phi^3)$. For instance, 
\begin{equation}
    iS_{\phi^3} 
    = \frac{ig_3}{3!m_p^3} \int_{p_i} \delta_{p_i}\left[ (p_1 \cdot p_2) p_3^2 + (p_1 \cdot p_2) p_3^2 + (p_1 \cdot p_2) p_3^2 \right] \phi(p_1) \phi(p_2) \phi(p_3).
\end{equation}

%\section{Horndeskian Correction to the Gravitational Radiation}

%\bibliographystyle{plain}
\bibliography{binary}

\end{document}